\documentstyle[12pt,epsf]{artmrs}
\newcommand{\beq}{\begin{equation}}
\newcommand{\eeq}{\end{equation}}
\begin{document}
\begin{center}
{\bf EFFECT OF SIZE DISPERSITY ON THE MELTING TRANSITION}
\end{center}
\noindent
M.R. SADR-LAHIJANY$^{1}$, P. RAY$^2$,S. T. HARRINGTON $^{1}$ 
AND H.E. STANLEY$^{1}$

\noindent
$^1$ Center for Polymer Studies and Department of Physics,
 Boston University, Boston, MA 02215 ,USA
\noindent
$^2$The Institute of Mathematical Sciences,
        CIT Campus, Madras - 600 113, India

\subsection*{\small ABSTRACT}
We present a molecular dynamics simulation study of the liquid-solid
transition in a two dimensional system consisting of particles of two
different sizes interacting via a truncated Lennard-Jones
potential. We work with equal number of particles of each kind and the  
dispersity $\Delta$ in the sizes of the particles is varied by 
changing the ratio of the particle sizes only.  
For the monodisperse case ($\Delta = 0$) and for small values of
$\Delta$, we find a first order liquid-solid transition on increasing the
volume fraction $\rho$ of the particles . As we increase $\Delta$, the
first-order transition coexistence region weakens gradually and completely
disappears at high dispersities around $\Delta = 0.10$ .  At these
values of dispersity the high density phase 
lacks long range translational order but
possesses orientational order with a large but finite correlation 
length. The consequences of this
effect of dispersity on the glass transition and on the
melting transition in general are discussed.

\subsection*{\small INTRODUCTION}

The liquid-solid transition in a system of densely-packed interacting
particles has attracted considerable attention in recent
years\cite{glaser}. Such a system undergoes a transition from a
disordered liquid phase to an ordered solid phase on increasing the
volume fraction of the particles.  It was further observed that
polydispersity in the sizes of the particles has a profound effect on
the transition. With increasing dispersion, the solid structure becomes
unstable and above a certain degree of dispersity the solid cannot form
at all\cite{Dickinson}. The consequence of this should be very important
from the experimental point of view since colloidal suspensions in
general do have particles of various sizes and show the liquid-solid
transition \cite{pieranski} and in the simulations of glass transition,
particles of different sizes are always considered \cite{kob}. Still,
the effect of size dispersity on the liquid-solid transition has not
received sufficient attention.  In this paper, we study the effect of
size-dispersity on the liquid-solid transition for interacting particles
in two dimensions.

The instability of the solid phase with increasing size dispersity is
not striking, as one would intuitively expect that a high dispersity
naturally destroys the crystal order needed to form a solid. But
molecular dynamics (MD) simulation studies in three dimensions
\cite{Dickinson}, and similar recent studies in two dimensions
\cite{Ito}, consistently show the gradual weakening of the first order
transition with increasing dispersity $\Delta$ and find the existence of
a critical value $\Delta_c$ where the line of first-order transitions
ends. At $\Delta_c$, one does not see the first order transition. This
prediction also arises from a study employing the density functional
theory\cite{Barret-Hansen} and simpler models of crystals\cite{Pusey}.  The phase diagram is remarkably similar to
the first order transitions ending in a critical point that one observes
in the temperature driven liquid-gas transition.  
We study the transition at and around $\Delta_c$ by carefully examining the
nature of the phases obtained at different densities and dispersities.

\subsection*{\small MODEL AND SIMULATION}

We report a molecular dynamics simulation study of a $50-50$ mixture of
Lennard-Jones (LJ) particles of two different sizes. The particles are
contained in a two dimensional box of linear size $L$ with periodic
boundary condition used on all walls. We have performed our simulations
for $N=400,2500$ and $10000$ particles. To each particle we assign a radius
proportional to its LJ diameter and define the density $\rho$ as
the ratio of the total area occupied by the particles to the total
area of the box.

The degree $\Delta$ of size dispersity is quantified by the relative
width of the bimodal particle size distribution function. Here we
present results for dispersities $\Delta=0$, $0.06$, $0.07$ and
$0.10$. All physical quantities are measured in reduced units in which
the average LJ diameter $\sigma$, LJ energy scale $\epsilon$ and the
mass of each particle are one. Our results are all collected from the
isothermal hyper-surface of the phase space with $kT=1.0$, where $k$ is
the Boltzmann constant. Pressure $P$ is computed using the virial
relation\cite{Allen-Tildesley}.

\subsection*{\small RESULTS}

\begin{figure}
\centerline{
        \epsfxsize=8.0cm
        \epsfbox{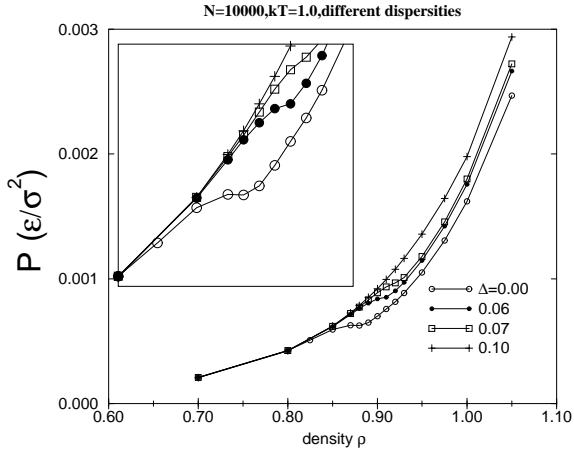}
        \vspace*{0.1cm}
        }
\caption{Pressure versus density plots for
different dispersities. The small dispersity curves show the first order
phase transition from the low density liquid phase to the high density 
2d-solid phase with the intermediate flat coexistence region.  
The inset is a blow
up of the coexistence regions at different dispersities.}
\label{figprho}
\end{figure}

Fig.~\ref{figprho} shows the $P-\rho$ diagrams for different
dispersities $\Delta$.  For small $\Delta$, we observe the flat 
coexistence region which is a characteristic feature of first order 
phase transition.  
This region shrinks as $\Delta$ increases and near $\Delta=0.1$
disappears completely.  

\begin{figure}
\centerline
	{
\hbox  {
        \epsfxsize=8.0cm
        \epsfbox{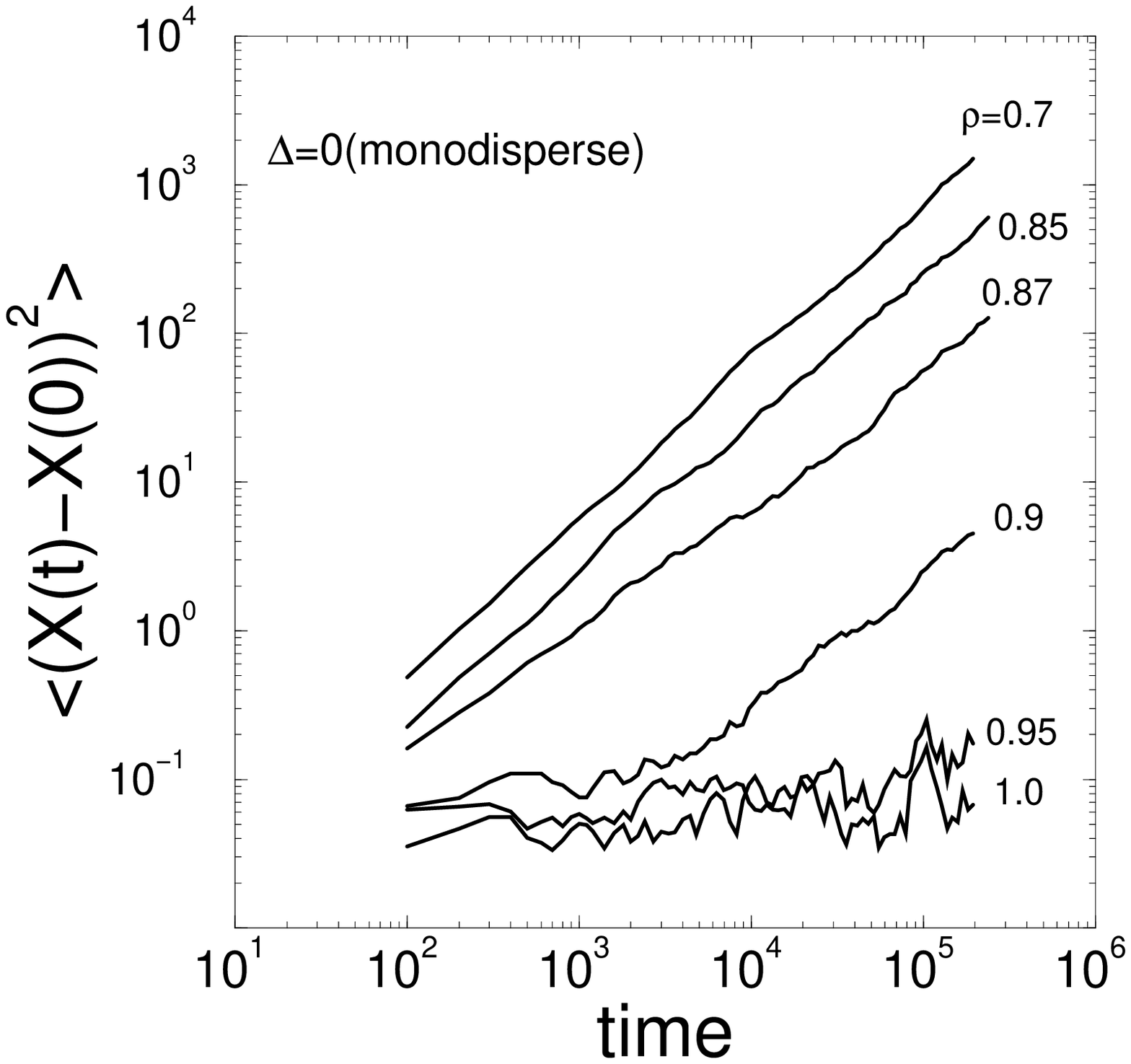}
        \hspace*{0.5cm}
        \epsfxsize=8.0cm
        \epsfbox{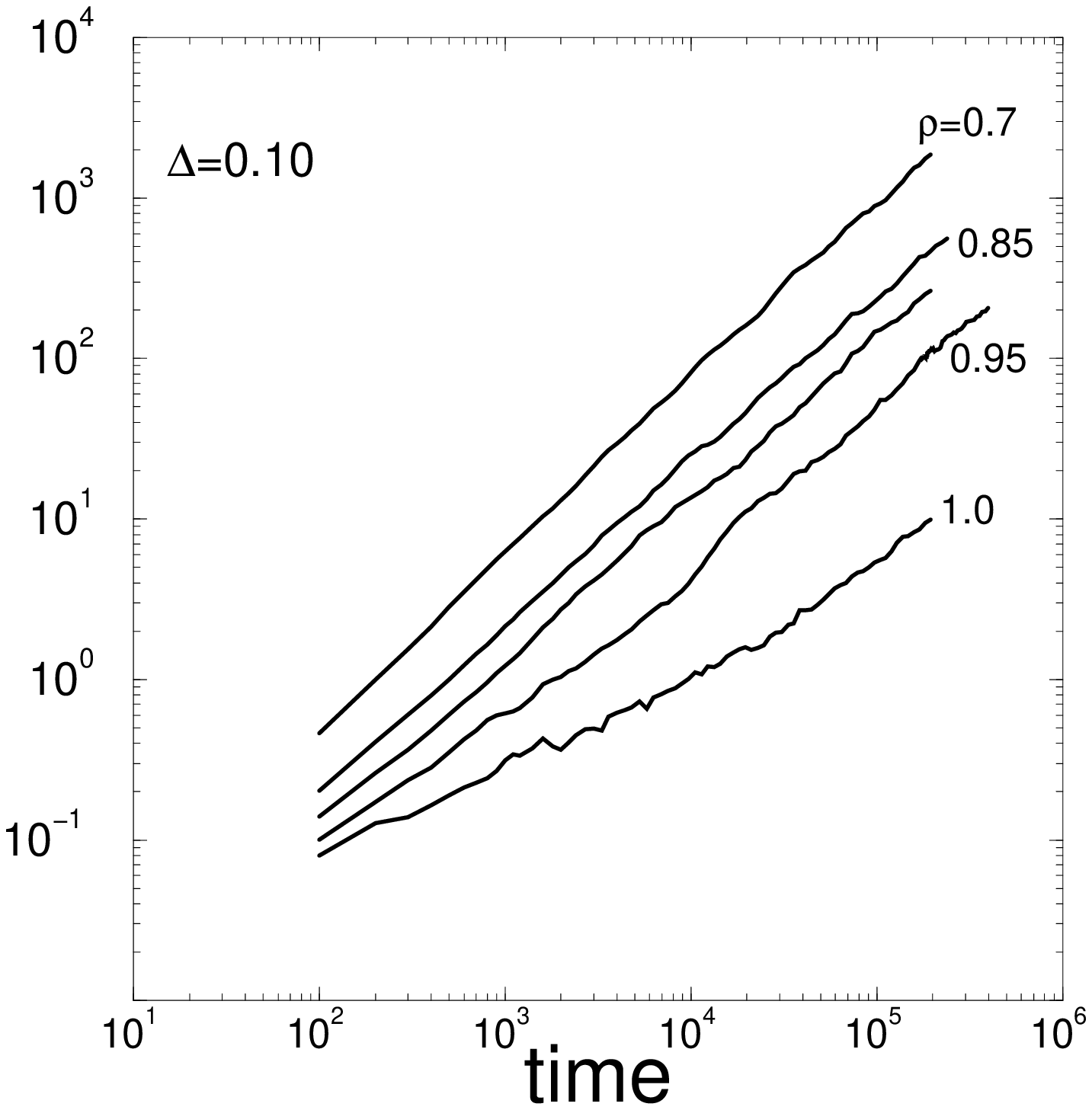}
	}
        }
\caption{Mean square displacement of the particles versus time at different 
densities $\rho$ (denoted at the end of the curve) for
 $\Delta=0$ and $\Delta=0.10$.}
\label{figmsds}
\end{figure}

%\paragraph{}
%Fig.~\ref{figconfigs} shows the configurations of the particles at
%equilibrium for different densities and dispersities, at different
%regions specified above. One can deduce the qualitative nature of the
%phases in the phases  from these figures. Our goal is to use some
%quantitative measures to determine and distinguish the characteristics
%of these phases.

\paragraph{}
Fig.~\ref{figmsds} shows the mean square displacement (MSD) of the particles
at different phases for $\Delta=0$ and 0.1.   
The plots for $\Delta=0$ essentially have three features: (i)
a late time diffusive regime for low densities (ii) a frozen regime
where the diffusion is very small for high densities (iii) a sudden
change in the MSD behavior between the two regimes on varying the
density. 
This jump is one of the characteristic features of a first-order
phase transition. We observe these features for other $\Delta$ values 
but with increasing $\Delta$, the magnitude of the jump in the MSD plot 
decreases and the system goes from liquid to solid regime rather   
continuously. On the other hand the MSD plot for $\Delta=0.10$ shows that at 
high dispersity, the system does not become solid even at $\rho=1.0$. 

\begin{figure}[hbt]
%\centerline{\bf PAIR DISTRIBUTION FUNCTION}
\centerline{
\hbox  {
        \epsfxsize=7.5cm
        \epsfbox{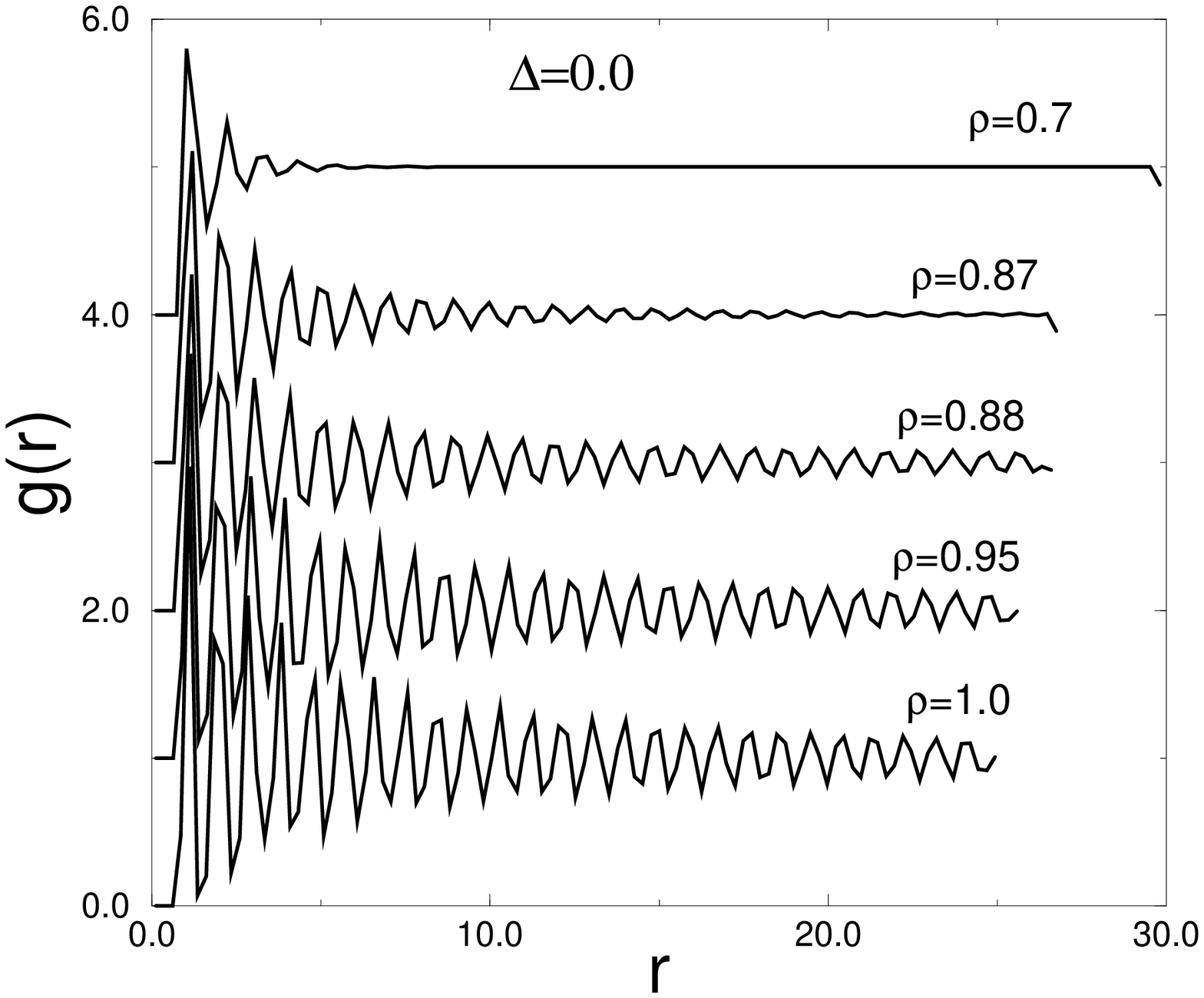}
        \hspace*{0.5cm}
        \epsfxsize=7.5cm
        \epsfbox{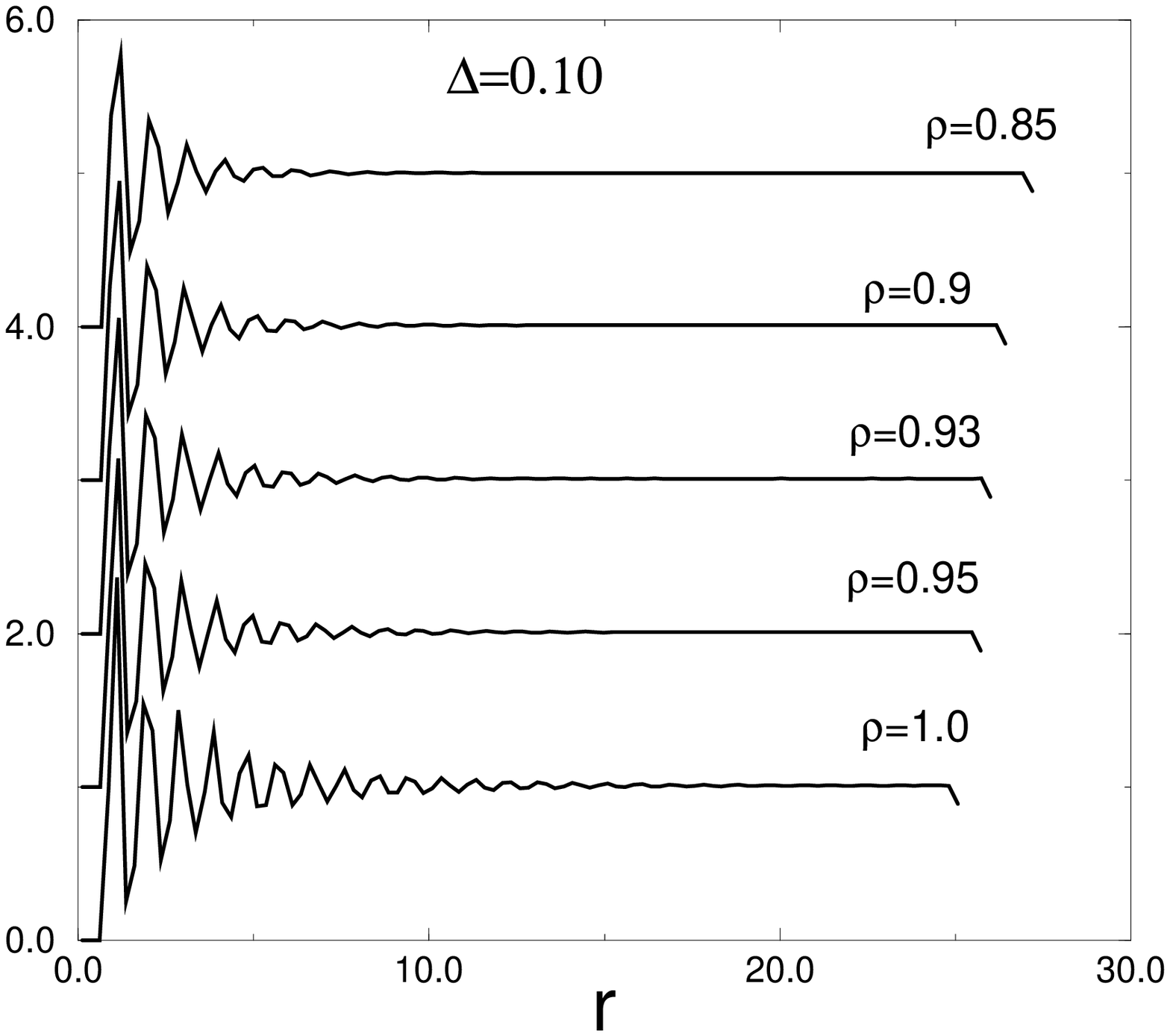}
	}
        }
\caption{Total pair distribution function for
two different dispersities, all graphs are centered around $g(r)=1.0$
but are shifted to make comparison easier;
 In monodisperse systems, the low density plots show the characteristics
of liquid and the high density ones that of solid. 
 For $\Delta=0.10$ systems, even the highest density studied , $\rho=1.0$
does not show the quasi-long-range order which is characteristic of a $2d$
solid. }
\label{figgrs}
\end{figure}
%\paragraph{}
In order to study the translational symmetry at different phases, we measure
the total pair distribution function $g(r)$ 
(Fig.~\ref{figgrs}).  The form of $g(r)$ for the low density phase has a
liquid-like structure. On the other hand for $\Delta=0$, the solid 
phase at high densities shows a clear 2d  
solid-like structure with pronounced peaks, and deep dips which persist
up to long distances, with the amplitude of the peaks decreasing
slowly. This quasi-long-range translational symmetry is expected for
solids in two dimensions \cite{strandburg}. For large values of
disperstiy around 
$\Delta=0.10$ and high density the system does not show
the solid structure mentioned above and lacks translational order.

\begin{figure}[hbt]
\centerline
	{
\hbox  {
        \epsfxsize=7.5cm
        \epsfbox{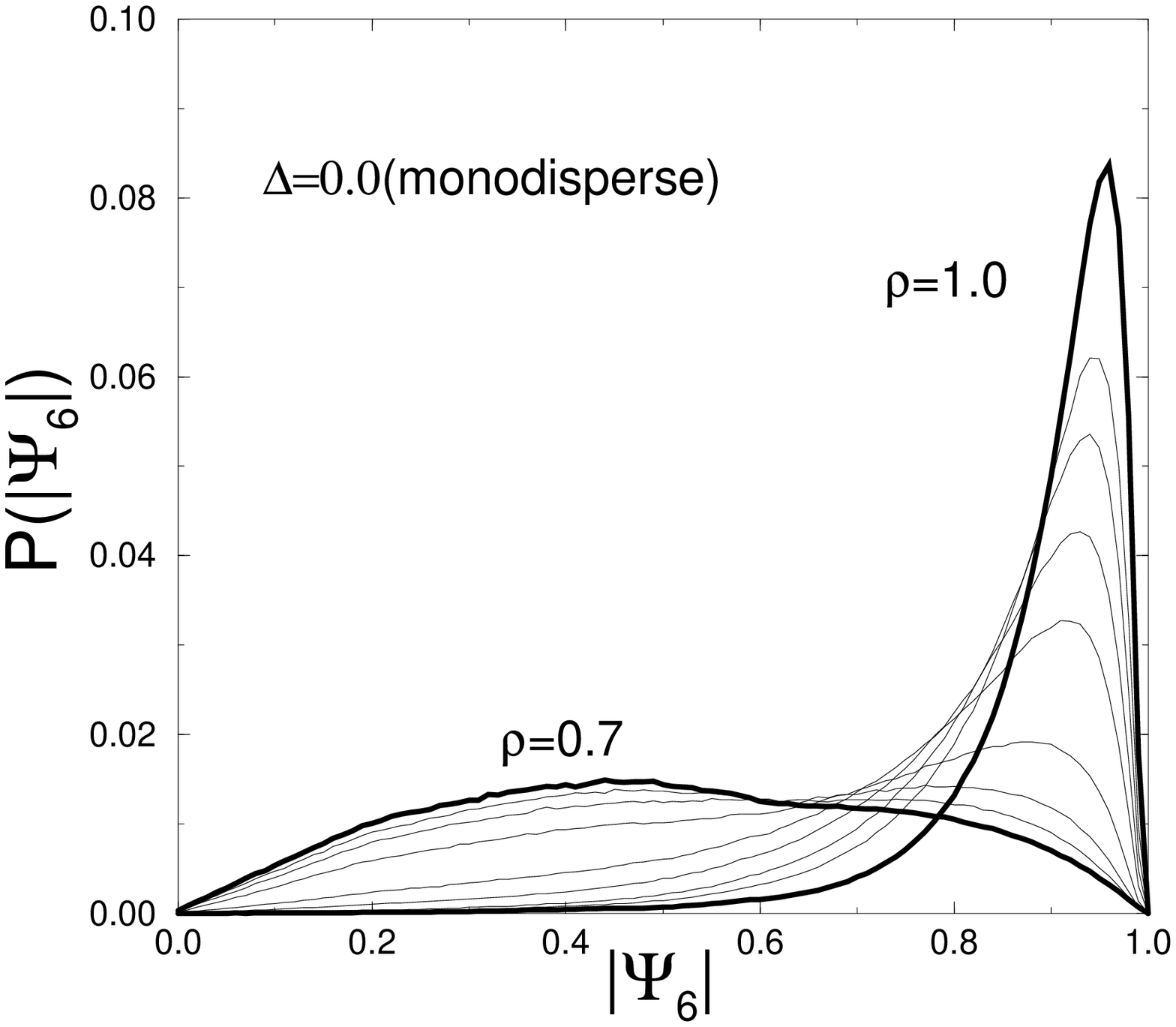}
        \hspace*{0.5cm}
        \epsfxsize=7.5cm
        \epsfbox{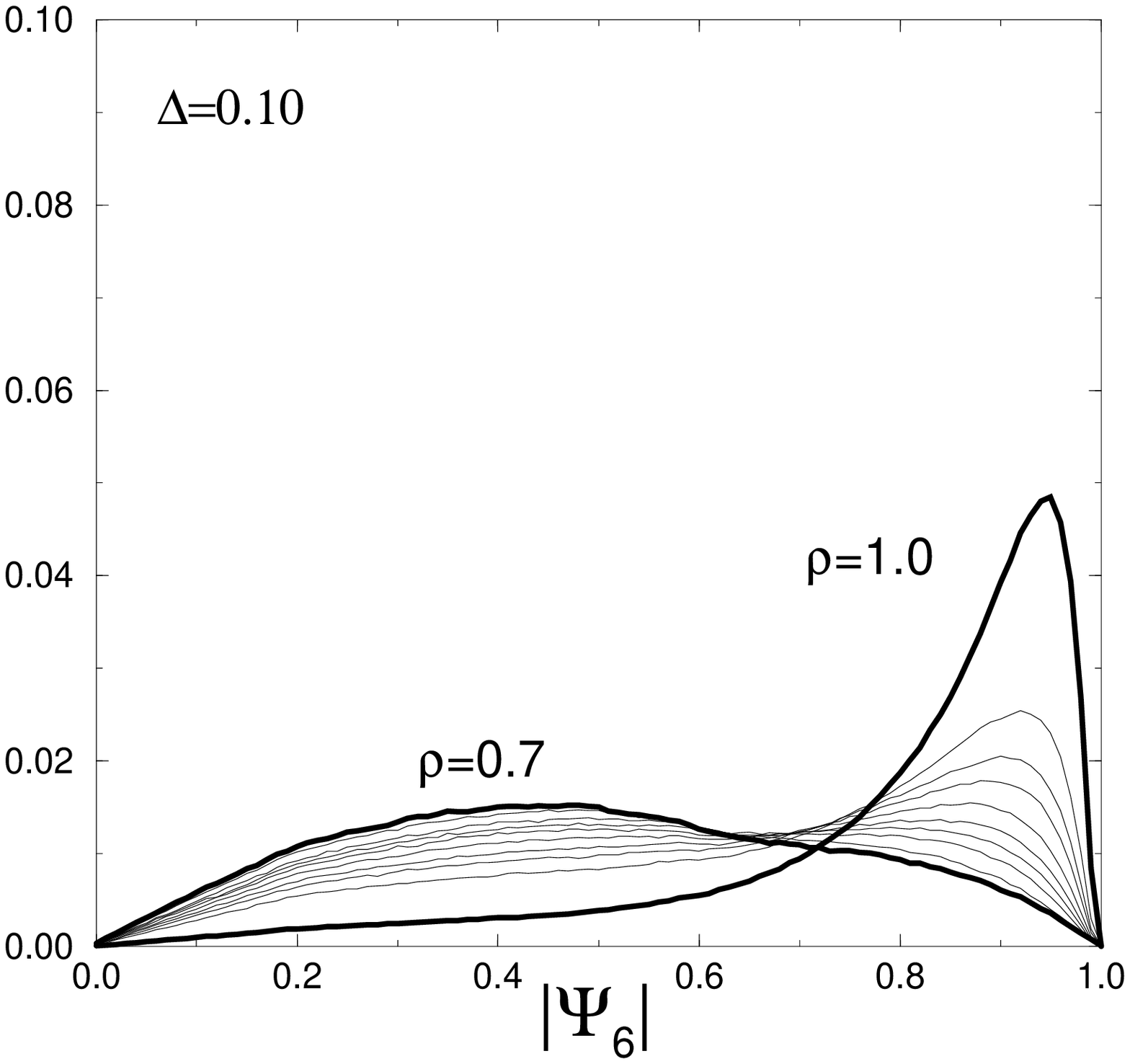}
	}
        }
\caption{$|\psi_6|$ distribution for two different dispersities;
 For monodisperse systems, $\rho=0.7$ plot belonging to the liquid phase
shows no orientational order, $\rho=0.875$ plot has a developing hump
along with a very fat tail indicating the coexistence of the solid and
liquid phases, $\rho=0.9$ to $\rho=1.0$ on the other hand have
pronounced peak near one and diminishing tails , indicating a triangular
lattice arrangement of the particles. 
On the other hand, for $\Delta=0.1$, even at $\rho=1.0$, we find a long 
tail in the distribution indicating the presence of many defects in the 
form of dislocations and disclinations and absence of true long range 
orientational order. }
\label{figpsi6}
\end{figure}

%\paragraph{}
Next, we study the orientational order of the phases by measuring the
hexagonal order parameter $\psi_6$ \cite{clark} which characterizes the
local bond orientational order around particles. The absolute value of
$\psi_6$ increases from a small positive value to one as the structure changes
from disorder to an ordered triangular lattice. We have plotted the
distribution of $|\psi_6|$ for all particles (see Fig.~\ref{figpsi6}).
The liquid phase has a flat distribution, thus not showing any local
orientational order \cite{clark}.  The solid phase shows a high degree of
local orientational order, since it forms a nearly perfect triangular
lattice. On the other hand for $\Delta=0.10$ the $\rho=1.0$ system 
shows a hump near unity but also a big tail extending down to zero. 
The hump confirms the existence of local 
hexagonal order which has also been observed in experiments on
bidisperse hard spheres\cite{Nelson}.  The presence of the long tail indicates that there are many
orientational defects in the system, as a result of size
dispersity. These defects are disclinations and appear as distorted
hexagonal or pentagonal and heptagonal neighboring particle arrangements 
around the particles.

\subsection*{\small CONCLUSION}

We find that a geometrical factor like the dispersity $\Delta$ in the
particle sizes has a similar effect on the liquid-solid transition as the
temperature (thermal energy) has on the liquid-gas transition. $\Delta$
weakens the first order transition from the liquid state to the solid
state driven by the volume fraction $\rho$ of the particles. This   
observation supports the earlier similar observations in different
systems like in elastic disk systems\cite{Ito} and in colloidal systems
\cite{Dickinson} with polydispersity in the sizes of the
particles. We further observe that at high values of $\Delta$, the 
system always remains at the fluid phase. This is the region where one 
can observe glass transition. Our study provides a quantitative 
measure of the size dispersity $\Delta$, which would be needed to 
observe the glass transition. It further indicates that there may not 
be any true phase transition (in the thermodynamic sense) in the process 
of glass transition.  We will provide detailed evidence for this
conclusion elsewhere\cite{unpublished} ( see also \cite{Dasgupta}).
However, much detailed study, specially on  
the effect of temperature is needed to say anything conclusively.  

\subsection*{\small ACKNOWLEDGMENTS}
We wish to thank L. Amaral, B. Kutnjak-Urbanc and W. Kob  
for useful discussions and remarks. The Center for Polymer studies is
supported by NSF.

%\begin{figure}
%\caption{Space configuration of the particles at equilibrium, for different
%densities and dispersities, belonging to
%different regions marked in the $P, \rho$ diagram;
%{\bf(a)} $L$ phase: $N=270,\Delta=0.0$ ($\rho\sim0.7$),
%{\bf(b)} $L$ phase: $N=270,\Delta=0.07$ ($\rho\sim0.7$),
%{\bf(c)} $C$ region: $N=350,\Delta=0.0$ ($\rho\sim0.875$),
%{\bf(d)} $S$ phase: $N=380,\Delta=0.0$ ($\rho\sim0.95$),
%{\bf(e)} $S$ phase: $N=380,\Delta=0.05$ ($\rho\sim0.95$),
%{\bf(f)} $S$ phase: $N=380,\Delta=0.07$ ($\rho\sim0.95$),
%{\bf(g)} $H$ phase: $N=380,\Delta=0.10$ ($\rho\sim0.95$)}
%\label{figconfigs}
%\end{figure}

\end{document}